\documentclass[review]{elsarticle}

\usepackage{lineno,hyperref}
\modulolinenumbers[5]
\usepackage{xcolor}

\newcommand{\etal}{et al.~}

\journal{Journal of \LaTeX\ Templates}

\begin{document}

\begin{frontmatter}

\title{Learning with hidden variables}

\author{Yasser Roudi}
\address{Kavli Institute \& Centre for Neural Computation,  NTNU, Trondheim, Norway}
\address{Nordita, Stockholm, Sweden}
\author{Graham Taylor}
\address{School of Engineering, University of Guelph, Guelph, Canada}

\begin{abstract}
Learning and inferring features that generate sensory input is a task continuously performed by cortex. In recent years, novel algorithms and learning rules have been proposed that allow neural network models to learn such features from natural images, written text, audio signals, etc. These networks usually involve deep architectures with many layers of hidden neurons. Here we review recent advancements in this area emphasizing, amongst other things, the processing of dynamical inputs by networks with hidden nodes and the role of single neuron models. These points and the questions they arise can provide conceptual advancements in understanding of learning in the cortex and the relationship between machine learning approaches to learning with hidden nodes and those in cortical circuits.
\end{abstract}
\begin{keyword}
statistical models, deep learning, dynamics
\end{keyword}

\end{frontmatter}

\section{Introduction} 

Learning the fundamental features that generate sensory signals and having the ability to infer these features once given the sensory input is a crucial aspect of information processing. It would not be far fetched to hypothesize that the organization of cortical circuitry is largely evolved for performing such tasks, or to put it slightly differently, that our evolutionary history has stored memories of such features in the connections that form a large part of our brains. But how can a neuronal network learn and extract features that cause the experiences of our sensory organs, in a supervised or unsupervised manner? Answering this question has been, and still is, a fundamental task of theoretical and experimental neuroscientists and those interested in artificial intelligence. And as one can imagine, it is a question that can lead, and in fact already has contributed, to deep insights about the organization of cortical microcircuitry and the way organisms understand their surrounding environment. 

\begin{figure}[h!]
\centering
\includegraphics[width=1 \textwidth]{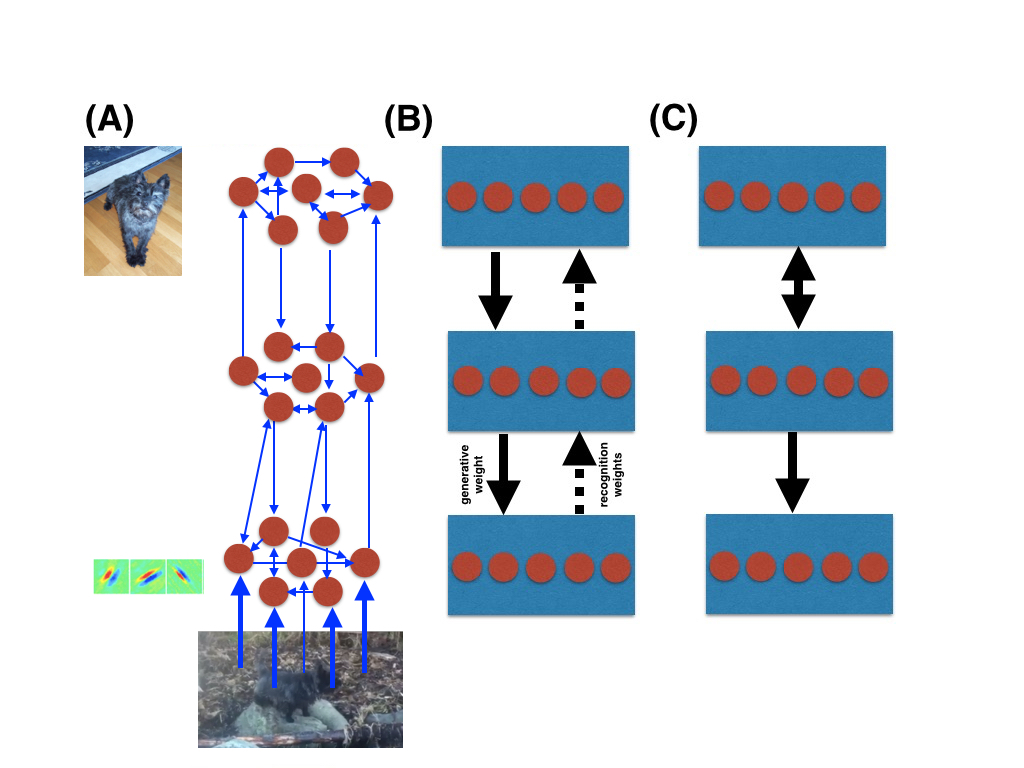}
\caption{{\bf (A)} The external world generates activity in the early sensory areas which then transfer it to higher cortical layers. Learning changes the connections between neurons such that the system learns the generative distribution of the observed images. This usually leads to layers gradually discovering features with various levels of complexity: simpler features such as edge detectors in the early sensory areas and more complex and abstract features, e.g. a dog, appearing  at the top level away from sensory organs. In this cartoon, the network does not necessarily represent a probabilistic graphical model, but more a stochastic dynamical system in which the connections reflect the anatomical connection between neurons. (B) The Helmholtz Machine is a graphical model which aims at modeling the function of the hierarchical cortical organization in a simplified probabilistic language. The full arrows indicates the set of generative weights while the dashed ones indicated the collection of recognition weights. Note that, although in general there is no reasons for not having connections between nodes in the same layer, and that this is certainly not the case in the brain, efficient learning considerations usually leads to models without intera-layer connections, and this is the case for the Helmholtz Machine and the related architectures. (C) The Deep Belief Network architecture. The double arrow between the last two layers indicate reciprocal and symmetric connections leading to an architecture called the Restricted Boltzmann Machine.}
 \label{Fig1}
\end{figure}

Over the past decades, it has become clear that the processing of sensory stimuli in the nervous system is performed in a hierarchical manner. Although hierarchical models of information processing leading to behaviour had already been suggested in the 50s \cite{mackay1956towards}, the experimental work of Hubel and Wiesel \cite{hubel1963shape} and the discovery of simple and complex cells inspired much activity in studying hierarchical models of cortical processing with various degrees of abstraction from high-level algorithmic models to neural network models. On the algorithmic side, there was the work pioneered by David Marr discussing the processing stages required for building a 3D perception of objects from the stimulus on the retina \cite{marr1978representation}. On the implementation side, mechanistic neural network models of visual pattern recognition were proposed \cite{fukushima1980neocognitron}. All these studies modeled sensory processing with a hierarchy of stages each capturing the necessarily basic features in the input useful for further analysis by higher levels.  This body of work, initially, did not include any probabilistic concepts: both the inputs and the computations performed at various stages were deterministic in nature, although some even earlier work did consider the probabilistic aspects of the sensory stimulus to be an important feature governing the design of the sensory information processing \cite{attneave1954some,barlow1961possible}. Over the years, more models emerged in which more emphasis was put on the probabilistic nature of stimuli. There has been a rise of models emphasizing the view that learning and inference of features of sensory input in a neuronal network can best be examined and answered by considering the hierarchy of processing stages as each performing probabilistic inference as depicted in the cartoon in Fig.\ \ref{Fig1} A. In this setting, one assumes that the sensory stimuli received by the sensory organs are generated through a probabilistic process as a combination of usually hidden causes which are then transmitted to the brain and need to be interpreted. By interpretation here, we mean the ability to use the sensory input to assign likelihoods to various combinations of the probabilistic causes that could have generated the input. The job of cortical processing layers and stages is to infer these probablistic hidden causes essentially through learning a generative model for the observed data. Building a successful model of cortical processing within this framework requires us to understand how in a hierarchy of cortical networks, the synaptic weights can be adjusted so that the probabilistic mapping from sensory input to underlying hidden causes that generated the input could be accomplished.

Although the initial ideas of hierarchical processing of sensory stimuli arose in the artificial intelligence community with the goal of ultimately understanding the operations of the nervous system, most of recent work and breakthroughs on the subject  have come from machine learning, and more recently, the deep learning community without much interest in understanding the brain. It is probably now time to re-examine these recent findings to see what are the implications for the nervous system. As we discuss below, some of the findings in the deep learning literature offer new directions of research and enquiry in neuroscience both at an algorithmic level and for biological implementation. In what follows, after a brief review of older relevant work on learning in neural networks, we describe a number of studies that have been responsible for the recent advances in learning in hierarchical stochastic models in the deep learning community. We then emphasize three issues which we think are of particular interest to neuroscience both for the insight they already offer and for the interesting research questions they raise: biological plausibility of recent learning algorithms, the role of single neuron input-output function in the success of learning, and learning from dynamic data.

\section{Learning hidden structure with hidden neurons}

A large body of early work in the neuroscience of learning hidden structures is focused on using networks with less emphasis on the probabilistic and unsupervised nature of the processing that what is common today and that we will discuss in the next section. In this previous work, one considers neuronal networks with potentially multiple hidden layers that progressively extract information about the input to perform an action, usually rewarded by a supervisory signal. The job of the learning rule was to adjust the synaptic weights such that the last stages of processing can e.g.~identify different input patterns as belonging to different categories as exemplified by a set of training data. In the 80s, the backpropagation algorithm \cite{rummelhart1986learning} provided significant excitement in the field as a potential mechanism for performing this task: the error between the present output and the desired output of a neuron is calculated starting from the final layer, propagated backwards in the network and used to adjust the weights. The algorithm, however, proved to be time consuming and easily stuck in unfavorable configurations of weights, let alone that its biological implementation was questionable. Consequently, there was a shift to other machine learning frameworks, such as support vector machines, with even more questionable biological implementations but ultimately useful for machine learning tasks. 

In addition to back-propagation and other supervised learning algorithms, unsupervised learning, in particular in probabilistic settings, also emerged in the 80s and the 90s. The Boltzmann Machine was proposed by Ackley, Hinton and Sejnowsky in 1985 \cite{Ackley85} as a model for learning to extract features from probabilistic sensory input with a particularly simple learning rule. Yet this also proved to be a very time consuming network, although faster approximations were also proposed. For instance, Kappen and Rodrigues \cite{Kappen98} and Tanaka \cite{Tanaka98} employed the susceptibility-response theorem from statistical physics combined with the mean-field solutions of the Ising model to devise algorithms for learning the weights in a fully observed Boltzmann Machine. Although the results were very promising, this line did not continue to the case with hidden nodes, and the use of mean-field methods (without any linear response considerations) have not been as successful \cite{galland1993limitations}.

Probably the first influential network architecture that was proposed for unsupervised learning of structure from data which used a computationally and biologically relevant learning algorithm is the Helmholtz Machine \cite{dayan1995helmholtz,hinton1995wake}. In the Helmholtz Machine, there is no supervised signal, and one takes the view that the stuff in the world are samples drawn from a probability distribution, the generative model, and the job of the nervous system is to learn this distribution via learning features to represent the data. This was carried out by a simplified model of Fig.\ \ref{Fig1}A, shown in Fig.\ \ref{Fig1}B which is comprised of a layer of observed neurons coupled to one or more layers of hidden neurons but without intra-layer connections. The inter-layer connections are divided into two sets of connections: a set going in the top-down direction called the generative weights, and a set going the opposite way called the recognition weights. As the name suggests, when the system is run by the generative weights, samples generated in the bottom visible layer are aimed to be samples from a distribution in the real world. To learn the generative weights, one uses the recognition mode to generate samples given data from the real world, and adjust the generative weights according to these samples. The opposite is done for learning the recognition weights. Although the Helmholtz machine is a powerful model both practically and conceptually, training it with many hidden layers is not easy. It can, however, be considered as a predecessor of the Deep Belief Network which proved to be a very successful and easily learnable model of complex data by employing a smart pre-training regime followed by fine tuning of the weights using the wake-sleep algorithm. We elaborate on this model below.

\section{Modern ways of learning in deep architectures}

Although the previous attempts in building learning rules for networks with hidden layers were all faced with some difficulties, more recent work on the subject, starting from the seminal work in 2006 \cite{hinton2006fast}, employed better understanding of the dynamics of networks with hidden nodes, combined with faster computers and larger data sets to successfully learn networks with hidden layers to extract hidden features in data. New learning algorithms have been proposed for using Deep Belief Networks and Deep Boltzmann Machines to extract information about the sensory signals in unsupervised, supervised, and semi-supervised learning tasks achieving unprecedented performance levels on industrially-relevant applications. After briefly reviewing these architectures and the learning rules used for training them we review some of the success stories most relevant to neuroscience in recent years. 

The introduction of Deep Belief Networks (DBNs) (Fig.\ \ref{Fig1}C) by Hinton \etal \cite{hinton2006fast} is widely viewed as the breakthrough that stimulated the modern view of \emph{Deep Learning}. Up to this point, classical multi-layer architectures such as deep neural networks were known to the machine learning community but were universally accepted as difficult to train in practice for all but the simplest of problems. Hinton and colleagues showed that a deep neural network could be grown incrementally, by composing single-layer models called Restricted Boltzmann Machines (RBMs). This simple symmetric neural network had been introduced many years earlier \cite{smolensky1986} but did not receive much attention until the early 2000's when the contrastive divergence learning algorithm was introduced \cite{Hinton2002} making approximate learning tractable.

To train a DBN, the individual RBMs were trained in sequence, each model learning to represent the one before it. Once this \emph{greedy} initialization strategy completed, the model could be \emph{fine-tuned} end-to-end. Labels could be introduced in this second stage of learning. By first discovering representations, learning became effective and efficient. This pre-training strategy has been empirically shown to lead to models with better generalization that seems to be more robust to random initializations \cite{erhan2010does}. While Hinton's group at the University of Toronto focused on RBMs as a basic building block, other groups proposed different unsupervised architectures. Bengio's group at the University of Montreal considered types of autoencoders, explained in more detail below \cite{Larochelle2007, Vincent2008}. LeCun's group at New York University (NYU) focused on sparsity as an alternative feature learning strategy \cite{Ranzato2007, Ranzato2008, Kavukcuoglu2008}. 

Once a Deep Belief Network has been composed by stacking, it is what is called a \emph{hybrid} directed-undirected model. That is because its topmost two layers which form an auto-associative memory are symmetrically connected like an RBM, but the lower layers are connected via directed links. Together, they form the \emph{generative} model. The DBN also contains a set of separate, bottom-up connections for fast approximate inference. They are not part of the generative model, but, similar to the Helmholtz machine, form a recognition model. It is this fast approximate inference which made DBNs very popular for several years, but it is also what limits them. It is well-known that the brain's wiring consists of feedforward, lateral and feedback connections. However, inference and generation in a DBN involves no feedback at all.

Salakhutdinov and Hinton \cite{salakhutdinov2009deep} showed how to train a related but different model called a Deep Boltzman Machine (DBM), in which feedback connections are important. In contrast to the hybrid DBN, the DBM is completely undirected. Like the DBN, its posterior distribution is intractable and must be approximated, however, the approximation is more powerful and more complicated than feed-forward recognition model of the DBN. Training the DBM involves a multi-step approach. It is first pre-trained, initialized from RBMs, much like the DBN. It is then generatively fine-tuned using a variational approximation to log-likelihood. Though the DBM has been shown to be useful in several applications, such as multi-modal learning \cite{srivastava2012multimodal} and one-shot learning with Bayesian priors \cite{salakhutdinov2013learning}, the standard approach to training a DBM is cumbersome. It requires training multiple models using different objective functions. Avoiding pre-training fails to learn a good model of the data. % Goodfellow \etal have recently proposed an alternative training strategy for DBMs dubbed ``joint multi-prediction'' training \cite{goodfellow2013multi}. The DBM is trained explicitly to predict any subset of variables given the complement of that subset. This work is among the first to show an effective method for joint training of all layers of a deep unsupervised learning model.   % using a mean-field variational approximation during the ``positive-phase'' where the model is driven by data, and a persistent chain stochastic approximation during the ``negative phase'' when the model is no longer clamped to the data. 
In response, current research is focused on the discovery of effective methods for
the joint training of all layers of a deep unsupervised model \cite{goodfellow2013multi}.

The last few years have seen a shift away from unsupervised learning back towards supervised learning, with an emphasis on an architecture called a convolutional network (convnet)\cite{lecun1998gradient}: a multi-layer neural architecture, inspired by Hubel and Wiesel's model of the visual cortex and early computational models such as Fukushima's Neocognitron \cite{fukushima1980neocognitron}. The architecture mainly consists of repeated stages of filtering (efficiently implemented by convolutions) and pooling, akin to the operation of simple and complex cells, respectively. 
Though proposed in the 1980's, the model has only reached widespread adoption in the last few years-- a result of larger datasets and faster computers \cite{krizhevsky2012imagenet}.
 % Until recently, the challenge of training convnets was commonly thought to be the fact that they would get stuck in local minima.  It is now understood that the local minima are nearly all similar, and the difficulty of training such networks results from saddle points surrounded by long plateaus along the error surface which can dramatically slow learning \cite{NIPS2014_5486}.
 % Larger datasets and faster computers --  specialized hardware accelerators called Graphics Processing Units (GPUs) have played a key role in the resurgence of convnets \cite{krizhevsky2012imagenet}.
 Architectural improvements, such as units that do not saturate and the use of normalization layers have also been important. Convnets have seen application to video 
\cite{Karpathy2014,Simonyan14b}, image captioning \cite{mao2014deep,xu2015show,vinyals2014show} and labeling entire scenes \cite{farabet2013learning}. The ImageNet dataset is commonly used as a means of supervised pre-training of filters to use on tasks for which large labeled datasets are not available \cite{razavian2014cnn}.

Despite the massive popularity of convnets, and the motivation for unsupervised pretraining tempered by success in pure supervised learning, unsupervised learning has not been abandoned. Following the dominance of RBMs in the mid-2000's, an unsupervised learning architecture known as the autoencoder has received increasing attention. An autoencoder is simply a feed-forward neural network which performs unsupervised learning by training it to reconstruct its input from a learned representation. Classical autoencoders use a bottleneck hidden layer, where the number of hidden units is less than the input, in order to avoid learning a trivial representation: simply a copy of the input which can be perfectly reconstructed. Recent work has explored alternative types of autoencoders which regularize, or restrict the hidden representation, for example by injecting noise: Denoising Autoencoders \cite{Vincent2008}, making the hidden units sparse \cite{kavukcuoglu2010fast}, or making the hidden units less sensitive to their input: Contracting Autoencoders \cite{Rifai2011}. Another body of work has generalized autoencoders to generative models \cite{bengio2014deep}. Generative Stochastic Networks define a generative model through a Markov Chain that injects noise at every stage of an iterated reconstruction. Like autoencoders, the model is easily trained with backpropagation yet is more capable of modeling multi-modal data distributions. A more recent line of work has revised the Helmholtz Machine architecture but improves on the inference procedure by correcting the gradient of the approximating posterior distribution. Several such deep variational networks \cite{Rezende-et-al-arxiv2014,Kingma+Welling-ICLR2014,Mnih+Gregor-ICML2014} were proposed concurrently by different groups and are considered the state-of-the-art in deep generative models. 

\section{Implications for cortical organization and plasticity}

How can the work in deep learning affect the work by theoretical and experimental neuroscientists aiming at understanding of the nervous system? Most of the work in deep learning so far has not had a direct dialogue with neuroscience and already some of the recent directions taken in deep learning are moving towards architectures and findings obviously away from our current understanding of the nervous system. In our opinion, however, there are already several findings and research questions which can and should inspire more work in neuroscience. We discuss three issues in this direction below.

{\bf Role of single neurons}. Due to the growing popularity of deep neural networks, many improvements have been proposed both at an architectural level, as well as in training methodology. However, two such discoveries, described in more details in Box 1, stand out in terms of their widespread adoption in modern deep neural networks both of which give insight into potential biological implementations: (1) the effect of the non-linearities used by individual neurons, (2) a regularization technique called dropout.

The observation that single neuron non-linearity plays an important role for the success of learning in deep architectures offers a potentially fundamental insight into the mechanisms of learning in the cortical hierarchy: that optimal synaptic weight changes, or whether such changes exist at all, is fundamentally tied to the single neuron properties. In retrospect this might sound like a trivial statement, but it is one that has been largely ignored. In the 80s, following the work of John Hopfield \cite{hopfield1982neural}, Daniel Amit \cite{amit1985spin} and others, statistical physicists showed interest in the studying of neural networks with binary neurons. The presence of a spin glass phase \cite{MezardParisiVirasoro} in which the network exhibited a large number of local minima, none of which related to the stored memories, was noticed and was probably an attractive ground for work by statistical physicists. Already then, however, it was noted that this phase is reduced in size in neuronal networks with sigmoid or piecewise-linear transfer functions \cite{kuhn1991statistical} and largely suppressed in networks of threshold-linear rectifiers \cite{treves1991spin}, and deemed more biologically plausible. This, intuitively, arises because the maximum slope of the transfer function acts as an effective inverse temperature meaning that the network can settle into a spin glass phase (usually a low temperature phase) at possibly higher temperatures for e.g.~rectifier neurons. The presumed heavily rugged energy landscape of the system in the spin glass phase, in turn, implies that the energy landscape of networks with graded response is much less rugged than the binary counterparts. We note that since these studies were concerned with models of the retrieval of already stored memories, the implications of these findings for learning in neuronal networks were not explored further. It is probably now time to take this issue up again and use the techniques of statistical physics to understand what single neuron properties and their relationship to the frustrated phases mean for learning in deep cortical models.

% could add that several variants have since been proposed as well as a number of papers giving a more theoretical way of explaining why the technique works

%\red{
%\begin{itemize}
%\item other learning algorithms for RBMs (persistent CD, score matching, MPF)
%\item autoencoders and regularized autoencoders
%\item unsupervised learning still continues on, recent work on variational autoencoders \cite{kingma2013auto} (relation to Helmholtz machines), adversarial generative training \cite{goodfellow2014generative}, backprop-free autoencoders %\cite{lee2014target} and more biologically plausable learning algorithms \cite{bengio2015towards}
%\end{itemize}
%}

{\bf Learning kinetics with hidden nodes.} Sensory signals are rarely static in real life and the learning of relevant features has to be done by taking these dynamics into account. Although, the overwhelming majority of the works we discussed above, both by neuroscientists and machine learning researchers, are focused on static data, as we describe below, more recent work has been focused on learning models in which these kinetic aspects are taken into account, some resulting in relatively simple learning algorithms. Given the dynamic nature of inputs to the brain, deep dynamic architectures should offer another interesting area of research for understanding learning in the cortical circuits.

Shortly after the discovery of DBNs, Restricted Boltzmann Machines were extended to capture temporal dependencies. The key insight was to condition both the observed and latent variables on fixed-length windows of the past instances of these variables. However, exact inference in such models is extremely hard and therefore it is computationally intractable to learn such models exactly. Sutskever and Hinton \cite{Sutskever2007a}, focusing on modeling synthetic video, proposed an approximate training strategy. Taylor et al. \cite{taylor2006modeling}, focusing on modeling human motion capture, proposed an architecture that avoided explicit temporal connections among latent variables. Sutskever \etal later proposed a third option which was a type of recurrent RBM architecture for which exact inference straightforward while maintaining full connections among latent variables \cite{sutskever2009recurrent}. Temporal variants of RBMs have been applied to 3-d tracking of people in video \cite{taylor2010dynamical}, autotagging music \cite{mandel2011contextual}, and facial expression transfer \cite{zeiler2011facial}.

Dynamical generative models are the same mathematical objects as kinetic models used in statistical physics to describe the dynamics of complex systems, e.g. spin glasses, particles in random media etc. In particular, the kinetic Ising model studied in spin glass physics is essentially a generalized linear model with a logit link function. It can also be thought as a system whose dynamics samples the state space of a belief network. In recent years, the problem of learning the weights in a kinetic Ising model given samples as well as predicting the statistics of the samples given the weights has been a very active area of research in statistical physics \cite{Roudi11,Zeng11,Mezard11}, leading several new approximate techniques to solve both the learning and inference problems either analytically or through very efficient computational methods. Most recently, some people have focused on the problem of learning and inference in kinetic Ising models with hidden nodes \cite{dunn2013learning,HertzandTyrcha2014,Bachschmid-RomanoandOpper2014,Battistin2015}. In particular, Battistin et al \cite{Battistin2015} have studied a model in which hidden factors are conditionally independent of each other given the observed ones, and have shown that both learning the weights and inference of the hidden states can be efficiently done using a message passing algorithm. For the kinetic Ising model, the Bayes optimal performance of the inference of the hidden spin configurations can be calculated using methods from statistical mechanics \cite{Bachschmid-RomanoandOpper2014}, something that would be interesting to do also for the equilibrium models. 

The last three years have seen an explosion of activity studying recurrent neural networks (RNNs), a generalization of feedforward neural networks which can map sequences to sequences. Training RNNs using backpropagation through time can be difficult, and was thought up until recently to be hopeless due to vanishing and exploding gradients \cite{hochreiter2001gradient}. Recent advances in optimization \cite{martens2011learning}, gated (multiplicative) units in the architecture \cite{hochreiter1997long,chung2014empirical} careful initialization \cite{sutskever2013importance,le2015simple} and tricks such as gradient clipping \cite{pascanu2012difficulty} have led to impressive results in modeling speech, handwriting and language \cite{graves2013speech,  graves2013generating,sutskever2014sequence,mao2014deep,vinyals2014show,xu2015show,mikolov2014learning,hannun2014deepspeech}. O'Reilly  and Frank \cite{o2006making} proposed a biologically-based algorithm that is similar to  the popular Long Short-Term Memory \cite{hochreiter1997long} both in  motivation and computation. The model, PBWM, implements a flexible working  memory with an adaptive gating mechanism, inspired by the  interaction between the prefrontal cortex and basal ganglia. Trained  by reinforcement learning mechanisms rather than backpropagation, it  performs comparably to LSTMs and other recurrent architectures on benchmark working memory tasks.

{\bf Biological Implications.} Learning rules currently used in deep learning are not biologically plausible. But is it possible, maybe by sacrificing a degree of performance, to derive learning rules that can be implemented in biological neuronal networks? This question should offer an interesting avenue of research and some ideas have already been suggested.

In a recent study, Bengio et al.~\cite{bengio2015towards} suggest that deep learning is possible using biologically plausible synaptic learning rules (e.g.~STDP) by performing approximate expectation-maximization (EM) like algorithms. What sets this work apart from other papers from the deep learning community is that, instead of starting from an algorithmic view and arguing for biological plausibility, it starts from the main learning rule observed in biological synapses (STDP) and reverse-engineers some objective function that the phenomenon could improve. The main finding is that the objective comes from a variational bound on the log likelihood of the data which leads to a variational EM learning algorithm. A further contribution of this paper is in connecting variational EM with noise injection, to training a denoising auto-encoder over both visible and latent variables. The architecture considered is a deep denoising autoencoder trained with Target Propagation \cite{lee2014target}, an alternative to backpropagation that computes targets rather than gradients at each layer. Compared to Boltzmann machines which are perhaps the best-known biologically plausible method for learning deep architectures, this technique completely avoids the need to avoid representative samples from the stationary distribution using (expensive) MCMC, which is both a practical advantage and more satisfactory from the point of biological plausibility.

% Overall, the contribution is that it is the first machine learning-based interpretation of STDP. 
 
In another recent study, Pavel and Miller \cite{Sountsov2015} implemented the learning rules for a Helmoltz Machine in a network of spiking neurons. The Helmholtz Machine is particularly suitable for neural implementation because its learning rules are indeed local. To implement the synaptic learning rule for the Helmholtz Machine, the delta rule, Pavel and Miller design a neural microcircuit of spiking neurons which act as the nodes of a Helmholtz Machine. Upon receiving the inputs from other units or from the external world, the different pools in each microcircuit follow a dynamics that leads to the required synaptic changes between the nodes as prescribed by the delta rule. This idea of considering local microcircuits for the implementation of learning rules is one that can potentially be exploited further, maybe even for the implementation of back-propagation like algorithms: one criticism agains the back-propagation is that the error signals should travel backwards along the axons. However, if individual nodes are not single neurons but microcircuits with both afferent and efferent connections to and from the neurons from the next layer and an internal mechanism to combine these signals for changing the effective connectivity between nodes such a problem may be avoided. 
 
%\red{Are these algorithms biologically plausible, and if not, how can we make them plausible? in kinetic models, can we learn features in an online way as the data comes in? can we learn Language this way? Learning very long-term %dependencies (e.g.~RNNs, LSTM resurgence). Learning with a memory (Memory networks \cite{weston2014memory} and Neural Turing Machines \cite{graves2014neural}).  Recent work showing that adversarial examples can ``poke %holes'' in neural networks \cite{Szegedy-ICLR2014}: focus on understanding and harnessing adversarial examples \cite{goodfellow2014explaining} for improved learning.}

\section{Conclusion}

The history of algorithm and architecture development for neuronal networks has drawn much inspiration from theoretical and experimental neuroscience.  Although the recent success of the field is remarkable, it has drifted away from biological plausibility towards application-driven abstract computational models. In this article, we have reviewed some of the most salient contributions to deep learning from the perspective of neurobiology. We argue that there are aspects of this line of work that should inspire our understanding of cortical circuits, and that more dialogue between machine learning practitioners and neuroscience should yield fruitful.

\section*{Box 1: Important modification with biological relevance}

As mentioned in the text, following the explosion of interest in deep neuronal networks, various new methods have been proposed to improve learning. Among these, the following have been of particular practical significance and appear to have clear implications for understanding biological neuronal networks.

{1. \bf Single neuron transfer function:} The first important observation that has lead to significant improvement in the learning is the use of non-saturating non-linearities in the individual neurons. In almost all cortical areas, single neurons operate far from their saturation regime, and this has lead to the use of non-saturating units (e.g.~threshold-linear units) in biological neuronal network models and the study of the effect of saturation or lack of it  in the function of such models\cite{treves1990graded,shriki2003rate,roudi2006localized}. In the deep learning side, it has been shown that non-saturating non-linearities, in particular, the rectified linear unit (ReLU) \cite{glorot2011deep} significantly improve convergence, both in unsupervised and supervised variants of deep neural networks. A generalized parametric form of the ReLU has lead to the first result which surpresses human-level performance on the ImageNet classification challenge \cite{he2015delving}. 

{\bf 2. Dropout:} The other notable discovery is the ``dropout'' regularization technique \cite{hinton2012improving}, which aims to prevent co-adaptation of hidden units by randomly shutting off a collection of units at each presentation of an input. This method has another interpretation of training multiple networks, one per example, but sharing all of their parameters. Obtaining a prediction amounts to taking the geometric mean of the probability distributions over all labels specified by the individual models. Dropout is partially motivated by a theory of the role of sex in evolution: the ability of a set of genes to be able to work well with another random set of genes makes them more robust. Similarly, each hidden unit in a dropout-trained network must learn to work with a randomly chosen set of other units. This encourages the hidden units to model useful features on their own rather than co-operating with other units.

\section*{References}
\bibliography{mybibfile}

\newpage
\section*{Highlighs}

\begin{itemize}
\item Work in machine learning has made learning in deep neuronal architectures possible.
\item Single neuron non-linearities make a strong impact on the success of learning.
\item Biological implementation of these learning rules are being suggested.
\item Dynamic nets with hidden nodes capture long-time correlations, relevant in biology
\end{itemize}

\section*{Highlighed Papers}
\begin{itemize}
\item Hinton et al.~\cite{hinton2006fast}{$^{\ast \ast}$}: This paper is widely thought to have launched the field of Deep Learning. It demonstrates how a generative model called a Deep Belief Network can be constructed by incrementally training and stacking Restricted Boltzmann Machines. It also introduces the idea of greedy layer-wise pre-training, an effective technique for initializing deep neural networks.
\item Salakhutdinov and Hinton \cite{salakhutdinov2009deep}{$^{\ast}$}: This paper demonstrates a tractable method for training Deep Boltzmann Machines (DBMs). Feedback connections play an important role in DBM inference, making them more consistent with biological architectures.
\item Pavel and Miller \cite{Sountsov2015}{$^{\ast}$}: This is the first paper to propose an implementation of the the Helmholtz Machine in a spiking neuronal network. Taking an intricate neuronal microcircuit as and individual node in a simple two-layer Helmholtz Machine, the authors show that the Helmholtz Machine learning rule can be implemented in the network.
\item Glorot and Bengio \cite{glorot2011deep}{$^{\ast \ast}$}: This paper shows that the use of non-saturating non-linear units has a very positive effect on the training of deep neuronal networks. Given the fact that cortical networks operate far from their saturating regime, this observation can be of great importance for understanding the success of learning from complex data in hierarchical cortical networks.   
\item Hinton et al.~\cite{hinton2012improving}{$^{\ast}$}: This paper introduces the regularization technique known as Dropout. Dropout and related methods are currently the most  effective means of regularizing large neural networks. The technique amounts to efficiently visiting a large number of related models at training time, while aggregating them to a single predictor at test time.
\item Taylor et al.~\cite{taylor2006modeling}{$^{\ast \ast}$}: This is the first paper to extend the Restricted Boltzmann Machine to unsupervised learning of sequences. It also demonstrates the advantages of using depth in sequence modeling.
\item Treves 1991 Phys Rev A \cite{treves1991spin}{$^{\ast}$} and Kuhn et al 1991 Phys Rev A. \cite{kuhn1991statistical}{$^{\ast}$}: Using spin glass techniques, these papers studied the effect of single neuron transfer functions on the retrieval properties of attractor networks. They found that the spin glass phase characterized by many local minima and metastable states is reduced in networks with graded response neurons. Although, this work is concerned with retrieval and not learning, the conclusion tallies well with recent findings in deep learning emphasizing the role of single neurons in making learning easier. Studying the relationship between these two results would thus be very interesting.
\end{itemize}
\end{document}